%% file: ISWCS_2014_CameraReady_v5.tex
\documentclass[conference,a4paper]{IEEEtran}

\usepackage{cite}
\usepackage{graphicx,color,epsfig,rotating}
\usepackage{amsfonts,amsmath,amssymb,bbm}
\usepackage{algorithm,algorithmic}
\usepackage{subfigure}
\usepackage{amsmath}
\usepackage{cite}
\usepackage{mdwtab}
\usepackage{placeins}
\usepackage{psfrag, graphicx}
\usepackage[latin1]{inputenc}
\usepackage{amssymb}
\usepackage{multirow}
\usepackage{mathrsfs}

\input{macros}

\newtheorem{defn}{Definition}
\newtheorem{theorem}{Theorem}

\begin{document}

\sloppy


\title{On the Average Performance of Caching and Coded Multicasting with Random Demands}


 \author{
   \IEEEauthorblockN{
     Mingyue Ji\IEEEauthorrefmark{1},
     Antonia M. Tulino\IEEEauthorrefmark{2},
     Jaime Llorca\IEEEauthorrefmark{2} and 
     Giuseppe Caire\IEEEauthorrefmark{1} 
    }
   \IEEEauthorblockA{
     \IEEEauthorrefmark{1}EE Department, University of Southern California.  
     Email: \{mingyuej, caire\}@usc.edu}
   \IEEEauthorblockA{
     \IEEEauthorrefmark{2}Alcatel Lucent, Bell labs, Holmdel, NJ, USA.   
     Email: \{a.tulino, jaime.llorca\}@alcatel-lucent.com}
 }



\maketitle

\begin{abstract}
For a network with one sender, $n$ receivers (users) and $m$ possible messages (files),
caching side information at the users allows to satisfy arbitrary simultaneous demands
by sending a common (multicast) coded message. In the 
worst-case demand setting, 
explicit deterministic and random caching strategies and explicit linear coding schemes have been shown to be order optimal. 
In this work, we consider the same scenario where the user demands are random i.i.d.,
according to a Zipf popularity distribution. In this case, we pose the problem in terms of the
minimum average number of equivalent message transmissions. 
We present a novel decentralized random caching placement and a coded delivery scheme 
which are shown to achieve order-optimal performance. 
As a matter of fact, this is the first order-optimal result for the caching and coded multicasting
problem in the case of random demands.
\end{abstract}

\section{Introduction}
\label{section: intro}
Content distribution services,  such as video on demand, catch-up TV and internet video streaming, are driving the exponential traffic 
growth experienced in today's networks \cite{cisco13}. 
Important features of this type of services are that user demands are highly predictable \cite{breslau1999web} and exhibit 
a high {\em asynchronous content reuse} \cite{ji2013throughput}. That is, while there exists a relatively small number of popular files that account for most of the traffic, users do not consume media in a synchronous way (unlike in live streaming or linear TV).  Typical users wish to access the desired content at arbitrary times, such that \emph{naive multicasting}~\footnote{Naive multicasting refers to the transmission of a common not-network-coded packet simultaneously 
overheard and decoded by multiple users, possibly at different quality levels, by using scalable coding and layered channel coding.} as implemented in 
Media Broadcasting Single Frequency Networks (MBSFN) \cite{mediaFLO}, is not useful.
Due to the increasing cost and scarcity of bandwidth resources, 
an emerging and promising approach for reducing network load consists of using 
{\em caching} directly at the wireless edge, e.g., at small-cell base stations or end user devices. 
The efficiency and throughput of different caching networks has been studied in 
recent works \cite{llorcatulino132, llorcatulino14, ji2013throughput, ji2013fundamental, maddah2012fundamental , maddah2013decentralized, niesen2013coded}. 

In \cite{llorcatulino132,llorcatulino14}, Llorca {\em et al.} presented a formulation for the general content distribution problem (CDP), 
where nodes in an arbitrary network are allowed to cache, forward, replicate, and code messages in order to deliver arbitrary user demands with minimum overall network cost.  The authors showed an equivalence between the CDP and the network coding problem over a so-called caching-demand augmented graph, which proved the polynomial solvability of the CDP under uniform demands, and the hardness of the CDP under arbitrary demands. 

In \cite{ji2013throughput, ji2013fundamental}, Ji {\em et al.} considered the one-hop Device-to-Device (D2D) wireless caching network, where user devices with limited storage capacity are allowed to communicate between each other under a simple protocol channel model \cite{gupta2000capacity}. By careful design of the caching configuration and the use of either coded \cite{ji2013fundamental} or uncoded \cite{ji2013throughput} devliery schemes, 
the throughput of the D2D caching network is shown analytically to scale as $\Theta\left(\max\left\{\frac{M}{m}, \frac{1}{n}\right\}\right)$,~\footnote{We will use the following standard {\em order} notation: given two functions $f$ and $g$, we say that: 1)  $f(n) = O\left(g(n)\right)$ if there exists a constant $c$ and an integer $N$ such that  $f(n)\leq cg(n)$ for $n>N$; 2) $f(n)=o\left(g(n)\right)$ if $\lim_{n \rightarrow \infty}\frac{f(n)}{g(n)} = 0$; 
3) $f(n) = \Omega\left(g(n)\right)$ if $g(n) = O\left(f(n)\right)$; 4) 
$f(n) = \omega\left(g(n)\right)$ if $g(n) = o\left(f(n)\right)$; 
5) $f(n) = \Theta\left(g(n)\right)$ if $f(n) = O\left(g(n)\right)$ and $g(n) = O\left(f(n)\right)$.} where $m$ is the total number of files, $n$ is the total number of users and $M$ is the per user storage capacity. Quantitative results of throughput of D2D caching networks under realistic propagation and topology models
are reported in  \cite{ji2013wireless}. 
These results show that, when $nM \gg m$, the throughput of a D2D caching network grows linearly with the per-user 
cache size $M$. From an operational viewpoint, this implies the remarkable fact that caching at the wireless edge
has the potential of turning Moore's law into a bandwidth multiplier: doubling the device memory capacity yields a two-fold increase in the 
per-user throughput.

A different approach is taken in \cite{maddah2012fundamental , maddah2013decentralized}, 
where Maddah-Ali {\em et al.} considered a single bottleneck caching network consisting of an omniscient transmitter (e.g., cellular base station) having access to the whole file library and serving all the $n$ users through a common shared link. 
Under the worst-case demand setting, 
constructed 
deterministic and random caching strategies along with 
network-coded multicast delivery schemes are shown to achieve the same throughput scaling law 
$\Theta\left(\max\left\{\frac{M}{m}, \frac{1}{n}\right\}\right)$ as in the D2D caching network, 
also shown to be within a bounded multiplicative factor from
an information theoretic cut-set bound, and therefore order-optimal.

Given that worst-case demands, in which each user requests a distinct file if possible, happen rarely in practice, 
we argue that it is more relevant to study the average performance when content requests follow a popularity distribution. With this motivation in mind, in this paper we study the {\em expected rate} (expected minimum number of equivalent file transmissions) 
in the single bottleneck caching network where the demands follow a Zipf popularity distribution, which is shown to be a good model for the measured popularity of video files \cite{breslau1999web}. 

We first propose a novel content distribution scheme, referred to as RAndom Popularity-based (RAP), that combines a random caching placement approach, characterized by a caching distribution that adapts to the content popularity and the system parameters, and a coded multicasting scheme based on {\em chromatic number} index coding \cite{blasiak2010index}. 
We derive the achievable expected rate in terms of an upper bound on the expected chromatic number of a certain random graph (see details later). 
We then propose a simpler scheme, referred to as Random Least-Frequently-Used (Random LFU), which approximates RAP and generalizes the well known LFU caching scheme.~\footnote{LFU discards the least frequently requested file upon the arrival of a new file to a full cache of size $M$ files. In the long run, this is equivalent to caching the $M$ most popular files.}  
In Random LFU,  each user just caches packets from the (carefully designed) $\widetilde m$ most popular files 
in a distributed and random manner. The delivery scheme is the same as RAP and hence based on chromatic number index coding. 
By introducing a novel scaling law approach 
for proving an information theoretic converse, we show the order-optimality of both RAP and Random LFU 
under a Zipf popularity distribution with parameter $\alpha$, where we distinguish the analysis for $0\leq\alpha<1$ and $\alpha > 1$. To the best of our knowledge, this is the first order-optimal result under this network model for nontrivial popularity distributions. 
In addition, our technique for proving the converse is not restricted to the Zipf distribution, 
such that it can be used to verify  expected rate order-optimality in other cases.  Finally, we verify our results by 
simulations and compare the performance of the proposed schemes with other state-of-the-art information theoretic schemes under different regimes of the system parameters. 

It is worthwhile noting that 
in a parallel and independent work \cite{niesen2013coded}, the same network model and expected rate minimization problem 
are considered. However, the scheme proposed in \cite{niesen2013coded} for an arbitrary popularity distribution does not guarantee order-optimality in general. 
Moreover, for some non-asymptotic regimes with finite $n, m$ and $M$, 
the expected rate of this scheme can be much worse than that of the order-optimal schemes proposed in this work. 
Due to space limitations, all proofs are omitted and can be found in \cite{mingyue2014}. 

\section{Network Model and Problem Formulation}
\label{section: network model}

We consider a network formed by a source node with access to a content library $\Fc=\{1,\cdots, m\}$  of files of size $F$ bits 
each, 
communicating to $n$ user nodes $\Uc=\{1,\cdots, n\}$ through a common 
broadcast shared link of finite capacity $C$. Without loss of generality, we can assume $C = F$ bits/unit time 
and measure the transmission rate of the scheme in units of  time necessary to deliver the requested messages to the users.  
Each user has a cache memory of size $MF$ bits (i.e., $M$ files). 
The channel between the content source and all the users follows a shared error-free deterministic model. 
Users requests files from the library in an independent and identically distributed way across users and over time, 
according to a popularity distribution $\qv=[q_f]_{f=1}^m$. 
The goal is to design a content distribution scheme (i.e., determine the information stored in the user caches and 
the multicast codeword to be sent to all users through the common link) such that all demands are satisfied with probability $1$ and the 
expected rate $\bar R(\qv)$ is minimized. We denote the minimum achievable expected rate by $\bar R^*(\qv)$ (which is also a function of $n,m,M$). 

Note that our problem is an instance of the coded content distribution problem (CDP) presented 
in \cite{llorcatulino132} for the specific case of the single bottleneck network. 
Moreover, for a given (uncoded) caching and demand configuration, finding the optimal transmission scheme 
in the single bottleneck network is equivalent to solving an index coding problem (ICP) \cite{birkkol98} 
with side information 
given by the chosen caching configuration. 
\section{Achievable Scheme}
\label{sec: Achievable Delivery Scheme}

In this section we present achievable schemes for the CDP in the single bottleneck network 
based on a (distributed) randomized popularity-based caching policy and a (centralized) index coding based 
delivery scheme. Order-optimality of the proposed schemes is shown in Section \ref{order opt}.

\subsection{Caching Placement Scheme}

We partition each file into $B$ equal-size packets, represented as symbols of $\FF_{2^{F/B}}$ for finite $F/B$. 
Let $\mathcal M$ and $\mathcal Q$ denote the  {\em packet level} 
caching and demand configurations, respectively, 
where $\mathcal M_{u,f}$ denotes the packets of file $f$ cached 
at node $u$, and $\mathcal Q_{u,f}$ denotes the packets of file $f$ requested by node $u$. We use Algorithm \ref{alg1} to let each user fill its cache independently (and therefore in a decentralized way) by knowing the 
caching distribution $\pv=[p_f]_{f=1}^{m}$, with $\sum_{f=1}^m p_f=1$ and $0 \leq p_f\leq1/M, \forall f$. 
This condition prevents from caching duplicated packets and violating capacity constraints. 
Notice that, while each user caches the same amount $p_f M B$ packets of file $f$, the randomized nature of the algorithm 
makes each user cache possibly different packets of the same file, which is key to maximize the amount of distinct packets 
of the same file collectively cached by network. It is immediate to observe that $p_f$ denotes the probability that a randomly (uniformly) chosen packet from the cache of any given user belongs to file $f$, and hence the reference to $\pv$ as the caching distribution.

\begin{algorithm}
\caption{Distributed Random Caching Algorithm} 
\label{alg1}
\begin{algorithmic}[1]
\REQUIRE $\pv=[p_f]_{f=1}^{m}$
\FORALL{$f \in \mathcal{F}$}
\STATE Each user $u$ caches a subset ($\mathcal{M}_{u,f}$) of $p_f M B$ distinct packets of file $f$ uniformly at random. 
\ENDFOR
\RETURN $\mathcal M = \{\mathcal{M}_{u,f}, u = 1, \cdots, n, f = 1, \cdots, m\}$.
\end{algorithmic}
\end{algorithm}

\subsection{Coded Multicast Delivery}
\label{sec: Transmission Scheme}

Our coded delivery scheme is based on chromatic number index coding \cite{blasiak2010index}.  
The (undirected) conflict graph $\mathcal H_{\mathcal M, \mathcal Q}$ is constructed as follows:
\begin{itemize}
\item Consider each packet requested by each user as a distinct vertex, i.e., if the same packet is requested 
by $N > 1$ users, it corresponds to $N$ distinct vertices.
\item Create an edge between vertices $v_1$ and $v_2$ if 1) they do not represent the same packet, 
and 2) $v_1$ is not available in the cache of the user requesting $v_2$, 
or $v_2$ is not available in the cache of the user requesting $v_1$.
\end{itemize}
Next, consider a minimum vertex coloring of the conflict graph $\mathcal H_{\mathcal M, \mathcal Q}$. 
The corresponding index coding scheme transmits the modulo sum of the packets (vertices in $\mathcal H_{\mathcal M, \mathcal Q}$) 
with the same color. Therefore, given $\mathcal M$ and $\mathcal Q$, the total number of transmissions in terms of packets is given by 
the chromatic number $\chi(\mathcal H_{\mathcal M, \mathcal Q})$. This achieves the transmission rate 
$\chi(\mathcal H_{\mathcal M, \mathcal Q})/B$.  

\subsection{Achievable Expected Rate}

Given $n,m,M$ and the popularity distribution $\qv$, our goal is to find the caching distribution $\pv$ that minimizes the expected 
rate $\bar R(\pv,\qv) \eqdef \EE[\chi(\Hsf_{\Msf, \Qsf})/B]$, where $\Hsf_{\Msf, \Qsf}$ denotes the random conflict graph which is a function of the random caching and demand configurations, $\Msf$ and $\Qsf$, respectively. 
The expectation is over i.i.d. (according to $\qv$) user demands. We can show that 
$\bar R(\pv,\qv)$ can be upper bounded by: 
\begin{align}
\label{eq:2}
\bar R(\pv,\qv) 
\leq \bar R^{\rm ub}(\pv,\qv) \eqdef \min\{\psi(\pv,\qv),\bar m\},
\end{align}
with high probability.~\footnote{The term "with high probability" means that $\lim_{F \rightarrow \infty} \mathbb{P}(\bar R(\pv,\qv) \leq \bar R^{\rm ub}(\pv,\qv)) = 1$. In the following, we first  let $F \rightarrow \infty$ and then let $n \rightarrow \infty$.} 
In (\ref{eq:2}), $\bar m=\sum_{f=1}^m \left(1 - \left(1 - q_f\right)^{n} \right)$ and 
\begin{eqnarray}
\label{eq: psi}
\psi(\pv,\qv) = \sum_{\ell=1}^n {n \choose \ell}  \sum_{f=1}^m \rho_{f,\ell} (1-p_f M)^{n-\ell+1} (p_f M)^{\ell-1},  
\label{eq:3}
\end{eqnarray}
where 
$\displaystyle \rho_{f,\ell} \eqdef \mathbb \PP(f = \arg\! \max_{j \in \Fc^\ell} \,\,\, (p_jM)^{\ell-1}(1-p_jM)^{n-\ell+1})$ 
denotes the probability that file $f$ is the file whose $p_f$ maximizes the term $\left((p_jM)^{\ell-1}(1-p_jM)^{n-\ell+1}\right)$ among $\Fc^\ell$ (the set of files requested by an arbitrary subset of users of size $\ell$).  
We denote the caching distribution that minimizes $\bar R^{\rm ub}(\pv,\qv)$ as $\pv^*$. In the following, we refer to the scheme that uses $\pv^*$ for caching according to Algorithm \ref{alg1} and chromatic number index coding for delivery as {\em Random Popularity-based} (RAP), with achievable rate $\bar R(\pv^*,\qv)$. 

Note that for any other caching distribution $\pv \neq \pv^*$, we have
\begin{align}
\label{optimization: 2}
& \bar R(\pv^{*},\qv) \leq \bar R^{\rm ub}(\pv^*,\qv) \leq \bar R^{\rm ub}(\pv,\qv), \quad \forall \pv\neq \pv^*.
\end{align}
Given that $\bar R^{\rm ub}(\pv^*,\qv)$ may not have an analytically tractable expression in general, we now present a simpler scheme that approximates RAP by using a caching distribution $\tilde \pv$ 
of the following form: 
\begin{align}
\label{ptilde}
& \tilde p_f = \frac{1}{\widetilde m}, \quad f \leq \widetilde m \notag\\
& \tilde p_f = 0, \quad f \geq \widetilde m + 1
\end{align}
where $\widetilde m \geq M$ is a function of $m$, $n$, $M$, $\qv$.

The form of $\tilde \pv$ is intuitive in the sense that each user just randomly caches packets (may not be the entire file) 
from the $\widetilde m$ most popular files by using Algorithm \ref{alg1}. 
In the case where $M$ is an integer and $\widetilde m = M$, this caching scheme coincides with LFU \cite{lee2001lrfu}. 
Therefore, we refer the proposed caching policy as {\em Random LFU}. 

There are two important aspects of Random LFU. 
First, given the caching placement, the chromatic number based index coding delivery scheme allows coding within the full set of requested packets, which is contrary to the coded delivery scheme considered in \cite{niesen2013coded}, where coding is only allowed within packets of specific file groups. 
Second, to guarantee the order optimality, the choice of $\widetilde m \in \{\lceil M \rceil, \cdots, m\}$ is essential and highly nontrivial. 
This makes the proposed caching scheme fundamentally different from the caching scheme of \cite{maddah2012fundamental}, 
which corresponds to the case $\widetilde m = m$ (referred as uniform caching). 
In fact, the choice of $\widetilde m \in \{\lceil M \rceil, \cdots, m\}$ balances the gains from local caching and coded multicasting as a function of the popularity settings and the system parameters. 
Moreover, notice that Random LFU  is only an approximation of RAP.~\footnote{The relationship between the actual rates of RAP and Random LFU, $\bar R(\pv^*,\qv)$ and $\bar R(\tilde \pv,\qv)$, is not known. However, it can be shown that $\bar R^{\rm ub}(\pv^*,\qv) \leq \bar R^{\rm ub}(\tilde \pv,\qv)$.} 
However, surprisingly, Random LFU is sufficient to guarantee order optimality under the class of Zipf popularity distributions \cite{breslau1999web}. 
In the following section, we will prove the order optimality of random LFU by using $\bar R^{\rm ub}(\tilde \pv,\qv)$. Note that, from (\ref{optimization: 2}),  this also implies the order optimality of RAP.


\section{Order optimality}
\label{order opt}
\begin{defn}
\label{def: order optimal}
A caching and delivery scheme is order optimal if its achievable expected rate $\bar R(\qv)$ satisfies: 
\begin{eqnarray}
\lim_{n \rightarrow \infty} \frac{\bar R(\qv)}{\bar R^*(\qv)} \leq \kappa,
\end{eqnarray}
where $\bar R^*(\qv)$ is the minimum achievable expected rate and $1 \leq \kappa < \infty$ is a finite constant independent of $m,n,M$.
\hfill $\lozenge$
\end{defn}

We will now show that Random LFU is order optimal for Zipf popularity distributions. Let the file popularity $\qv$ follow a Zipf distribution, given by 
$q_f = f^{-\alpha}/(\sum_{i=1}^m i^{-\alpha}), \, \forall f = \{1, \cdots, m\}$. 
We notice that 
the behavior of Zipf distribution 
is fundamentally different for the two regions of the Zipf parameter $0\leq\alpha < 1$ and $\alpha>1$,~\footnote{In this paper, we do not consider the case of $\alpha=1$. } which will be considered separately. 
We have:
\begin{theorem}
\label{theorem: gamma < 1}
When $0 \leq \alpha<1$, let $\widetilde{m} = \min\{$
$\left(n(1-\alpha)M/m\right)^{\frac{1}{\alpha}}m, m\}$, the expected rate of Random LFU  
is order optimal with high probability, 
and it is upper bounded by
\begin{align}
\label{eq: gamma < 1 1}
\bar R^{\rm ub}(\tilde \pv,\qv) &\leq \min\left\{\left(\frac{\widetilde m}{M}-1\right)\left(1 - \left(1 - \frac{M}{\widetilde m} \right)^{n\sum_{f=1}^{\widetilde m}q_f} \right) \right. \notag\\
&\left. + n\sum_{f=\widetilde m+1}^{m}q_f, m\right\} \leq \min\left\{\frac{m}{M}-1, n, m \right\}. 
\end{align}
with  $\tilde{\pv}$ given by \eqref{ptilde}. 
\hfill$\square$
\end{theorem}

Next, we consider the case of $\alpha > 1$. Depending on how the number of users $n$ scales with the library size $m$, 
we consider three different subcases: a)  $n = \omega\left(m^{\alpha}\right)$; b) $n = o\left(m^{\alpha}\right)$; c) 
$n = \Theta\left(m^{\alpha}\right)$.~\footnote{Due to space limitation, case c) is shown in \cite{mingyue2014} but it is omitted from the present paper.} 

\begin{theorem}
\label{theorem: gamma > 1 achievable 1}
When $\alpha>1$ and $n = \omega\left(m^{\alpha}\right)$, letting $\widetilde m=m$, the rate of Random LFU 
is order optimal with high probability and is upper bounded by
\be
\label{eq: gamma>1 1}
\bar R^{\rm ub}(\tilde \pv,\qv) \leq \min\left\{\frac{m}{M}-1 + o\left(\frac{m}{M}\right), m\right\}. 
\ee
with  $\tilde{\pv}$ given by \eqref{ptilde}. \hfill$\square$
\end{theorem}

\begin{theorem}
\label{theorem: gamma > 1 achievable 3}
When $\alpha>1$ and $n = o\left(m^{\alpha}\right)$, the rate of Random LFU 
is order optimal with high probability. Moreover, as $n \rightarrow \infty$, the values of $\widetilde m$ and the upper bounds of the achievable rate are 
given in Table \ref{table: table_1_1}  for different regimes of parameters $n,m,M$.~\footnote{Due to the limit of space, the case of $M = \Theta\left(n^{\frac{1}{\alpha-1}}\right)$ is shown in \cite{mingyue2014}.} 
\hfill$\square$
\end{theorem}

\begin{table}[ht]
\centerline{\includegraphics[width=8cm]{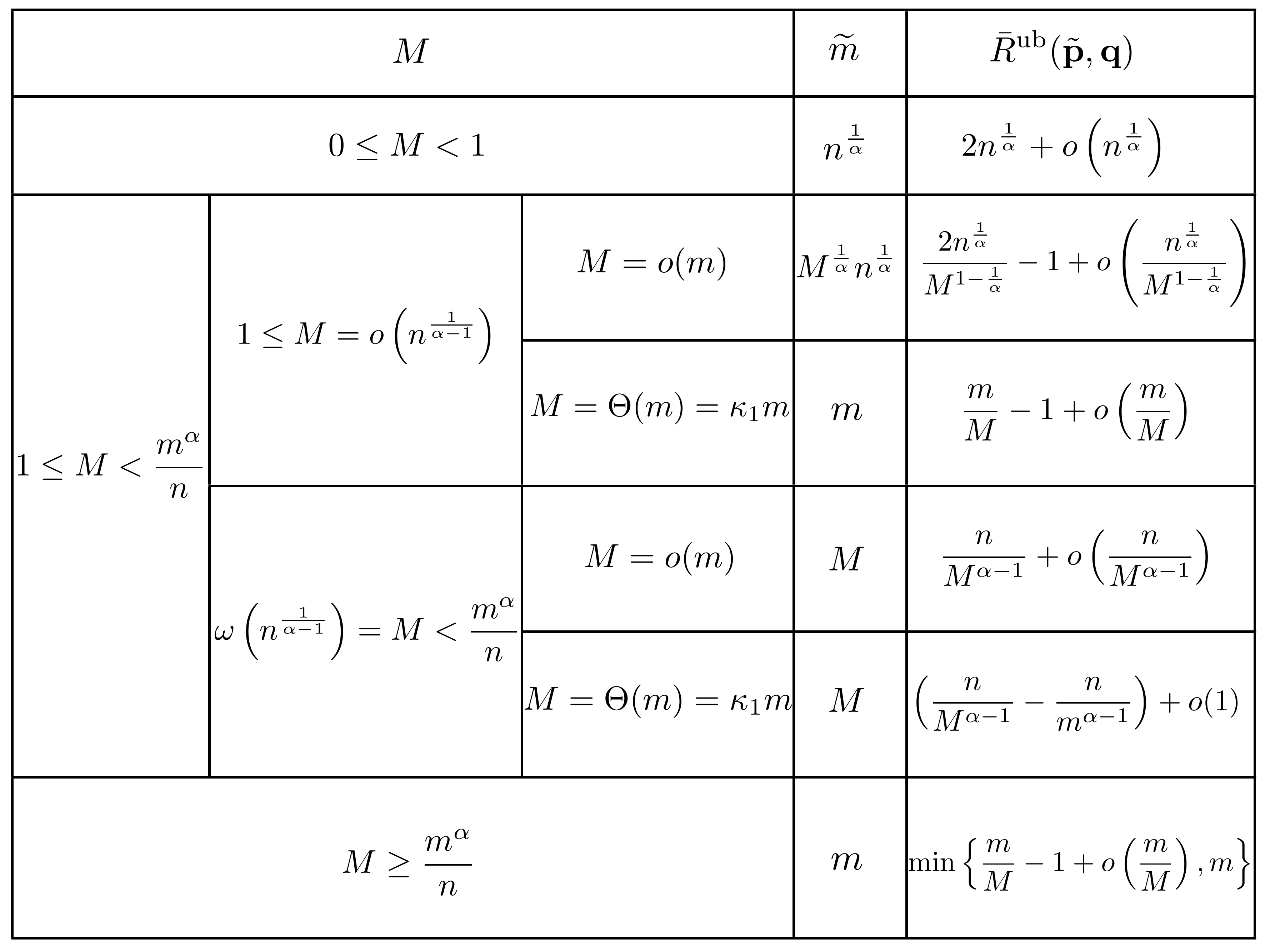}}
\caption{When $n = o\left(m^{\alpha}\right)$, this table shows $\widetilde m$ and the corresponding upper bound $\bar R^{\rm ub}(\tilde \pv,\qv)$ of the expected rate of Random LFU. In this table, $0<\kappa_1<1$ is a given positive constant.}
\label{table: table_1_1}
\end{table}

The converse of the expected rate, which is omitted due to its complex expression and can be found in \cite{mingyue2014}, is based on a modified cut-set bound. Qualitatively, we first find particular cuts of the caching-demand augmented graph under the guidance of $\widetilde m$ given in Theorem~\ref{theorem: gamma < 1}-\ref{theorem: gamma > 1 achievable 3} and then weight the lower bound of the rate obtained by each cut differently according to the demands distribution.

\section{Discussions}
\label{sec: Discussion}

\begin{figure}
\centering
\subfigure[]{
\centering \includegraphics[width=4.8cm, height=3.5cm]{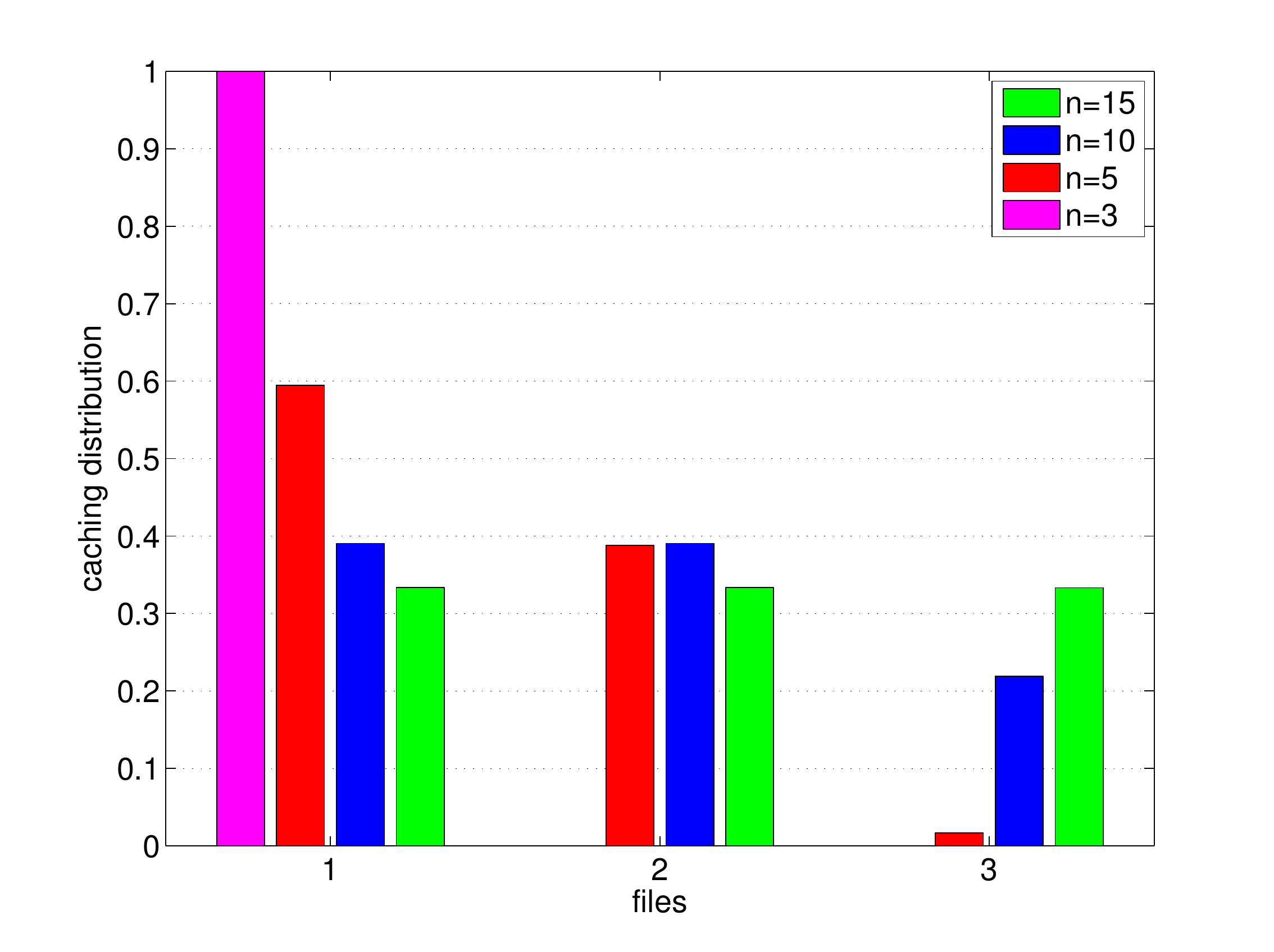}
\label{fig: p_opt_versus_files}
}
\vspace{-0.2cm}
\hspace{-0.7cm}
\subfigure[]{
\centering \includegraphics[width=4cm,height=3.4cm]{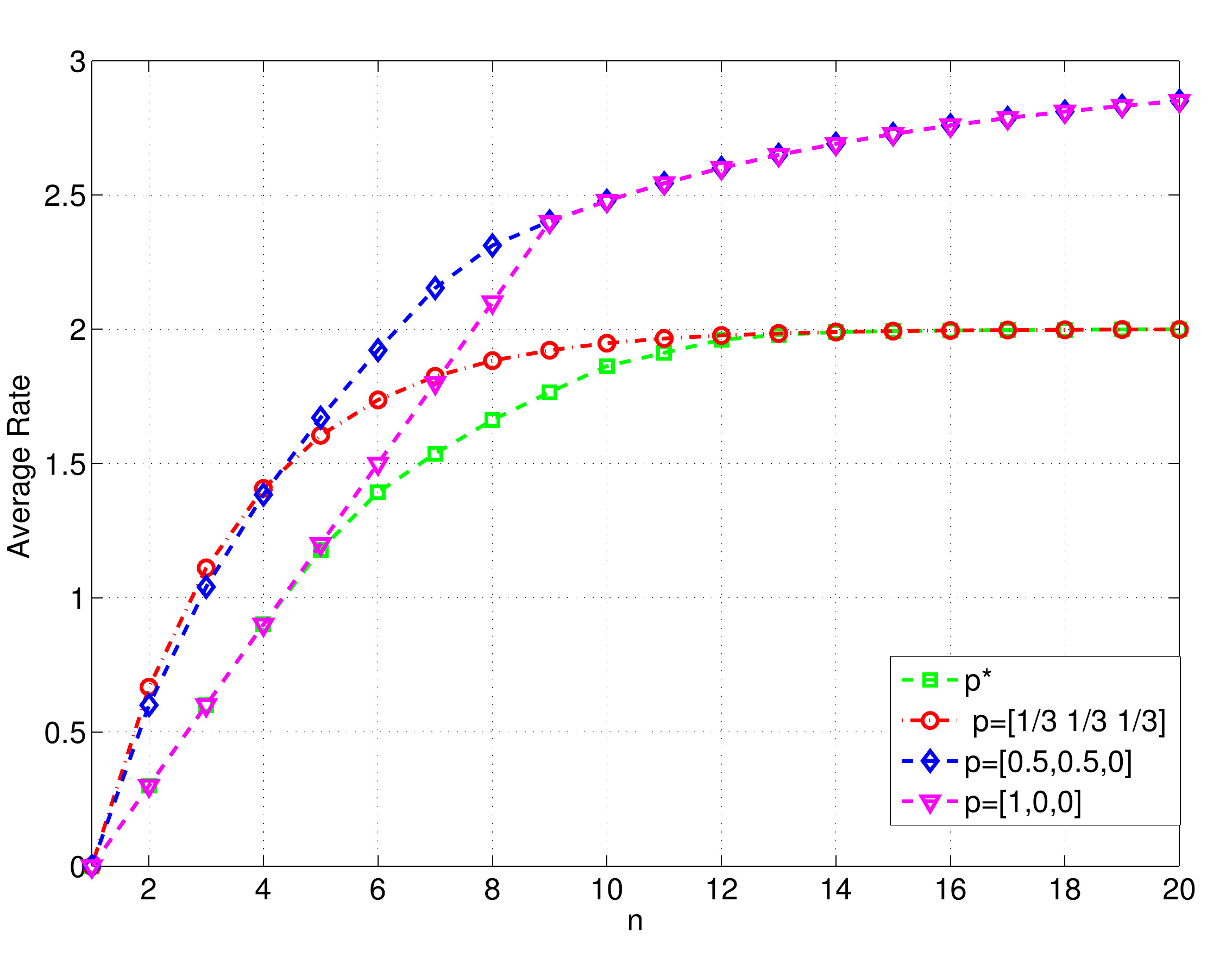}
\label{fig: Rate_optimized}
}
\vspace{-0.3cm}
\caption{Let $m=3$, $M=1$ and $n=3,5,10,15$ and the demands distribution be $\qv = [0.7,0.21, 0.09]$. a)~The caching distribution $\pv^*$. b)~The upper bound given by (\ref{eq:2}) of the expected rate by RAP. }
\vspace{-0.3cm}
\end{figure}

Fig.~\ref{fig: p_opt_versus_files} shows $\pv^*$ obtained by minimizing the upper bound $\bar R^{\rm ub}(\pv,\qv)$ given in (\ref{eq:2}) 
when $m=3$, $M=1$, 
$n=3,5,10,15$ and the demand distribution is $\qv = [0.7,0.21, 0.09]$.  
Observe how the caching distribution $\pv^*$, which does not necessarily coincide with $\qv$, adjusts according to the system parameters to balance the gains from local caching and coded multicasting. Note that $\pv^*$ goes from caching the most popular file (as in LFU) for $n=3$ to uniform caching for $n=15$. The corresponding achievable expected rate given by (\ref{eq:2}) is shown in Fig.~\ref{fig: Rate_optimized}, which confirms the average performance improvement when using the adaptive RAP distribution.
~\footnote{As mentioned before, the upper bound of the expected rate by using RAP with parameter $\pv$ is $ \bar R^{\rm ub}(\pv,\qv) = \min\{\psi(\pv,\qv),\bar m\}$, where $\bar m$ is obtained by just using the naive multicasting of the requested files without using caching. 
Hence, $\bar m$ can be improved by not sending the cached packets or using random linear coding \cite{maddah2013decentralized}. 
For example, in Fig.~\ref{fig: Rate_optimized}, when $\pv=[1,0,0]$, 
the server could just transmit the two uncached files such that rate is $2$, which is less than $\bar m$. 
However, this caching gain contributes at most by an additive constant.} 
Furthermore, from Theorems \ref{theorem: gamma < 1}-\ref{theorem: gamma > 1 achievable 3}, 
we can see that the caching placement characterized by $\widetilde \pv$ by using Random LFU has the same trends 
as $\pv^*$.  In particular, as $\alpha>1$, from Table~\ref{table: table_1_1} of Theorem~\ref{theorem: gamma > 1 achievable 3}, when 
$n = o\left(m^{\alpha-1}\right)$ and $\omega\left(n^{\frac{1}{\alpha-1}}\right)=M\leq m$ (approximately, this is the regime of $n$ small and $M$ is large), 
if the orders of $n$ and $m$ are fixed, let the order of $M$ increase, then the order optimal $\widetilde m$ varies from 
$n^{\frac{1}{\alpha}}$ via $M^{\frac{1}{\alpha}}n^{\frac{1}{\alpha}}$ to $M$, which means that the caching placement 
converges to LFU. Accordingly, the expected rate $\bar R^{\rm ub}(\tilde \pv,\qv)$ takes values from 
$\Theta\left(n^{\frac{1}{\alpha}}\right)$ via $\Theta\left(\frac{n^{\frac{1}{\alpha}}}{M^{1-\frac{1}{\alpha}}}\right)$ to $\Theta\left(\frac{n}{M^\alpha}\right)$, 
from which we can see that the effect of caching is also increasing in the sense that the average throughput (inversely proportional to the average rate) scales with $M$ from sub-linear to super-linear. 
When $\alpha > 1$ and $n = \omega\left(m^{\alpha-1}\right)$ (approximately, this is the regime where $n$ is large),  
if the orders $n$ and $m$ are fixed, let the order of $M$ increase, then the order optimal $\widetilde m$ varies from $n^{\frac{1}{\alpha}}$ via $M^{\frac{1}{\alpha}}n^{\frac{1}{\alpha}}$ to $m$, which means that the caching placement converges to uniform caching. 
Correspondingly, the scaling of the expected rate varies from $\Theta\left(n^{\frac{1}{\alpha}}\right)$, through  $\Theta\left(\frac{n^{\frac{1}{\alpha}}}{M^{1-\frac{1}{\alpha}}}\right)$ to $\Theta\left(\frac{m}{M}\right)$, where the effect of caching becomes more and more significant in the sense 
that the average throughput scales with $M$ from sub-linear to linear. 

\begin{figure*}[ht]
\centering
\subfigure[]{
\centering \includegraphics[width=4.7cm]{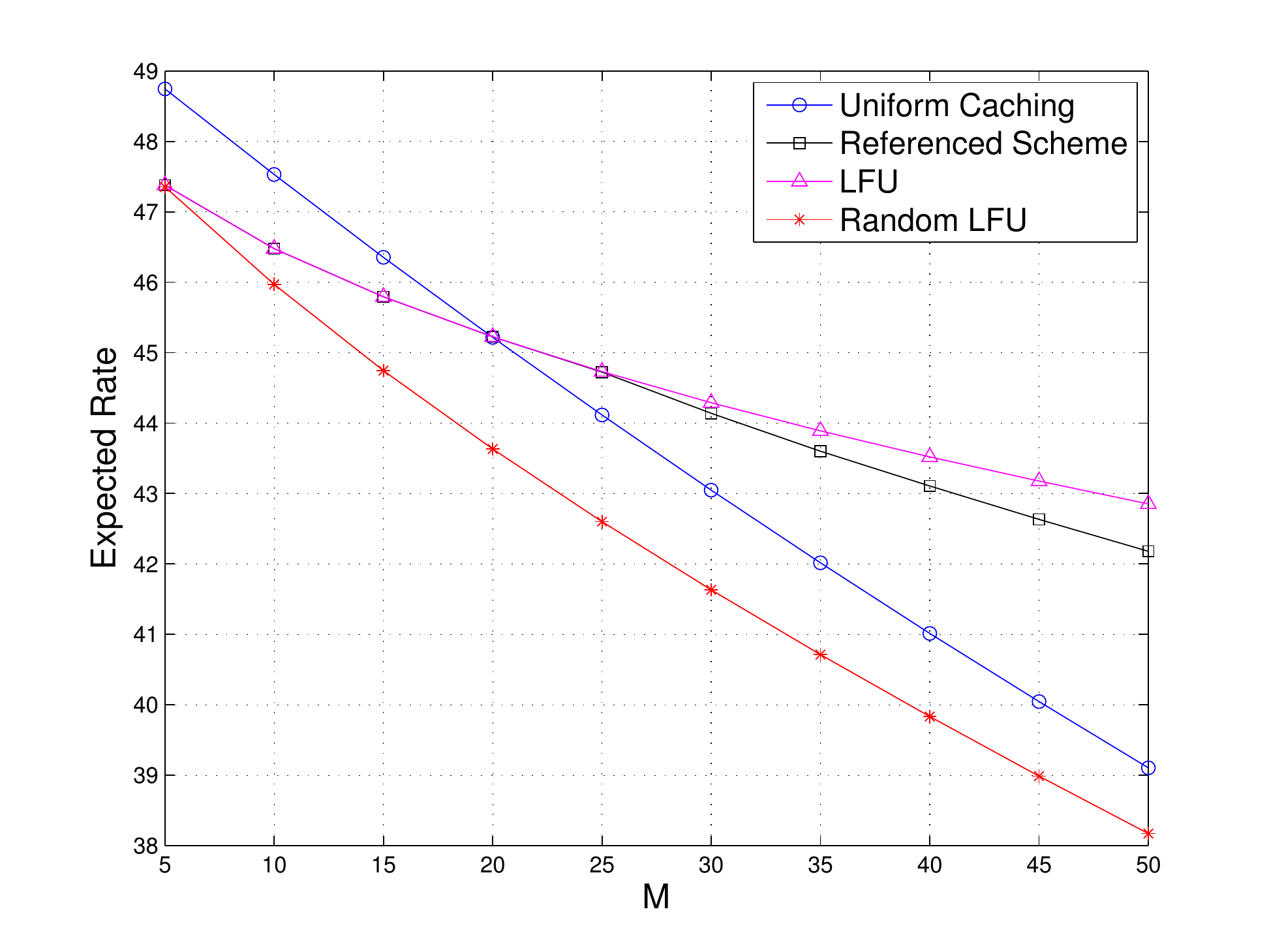}
\label{fig: }
}
\hspace{-0.8cm}
\subfigure[]{
\centering \includegraphics[width=4.7cm]{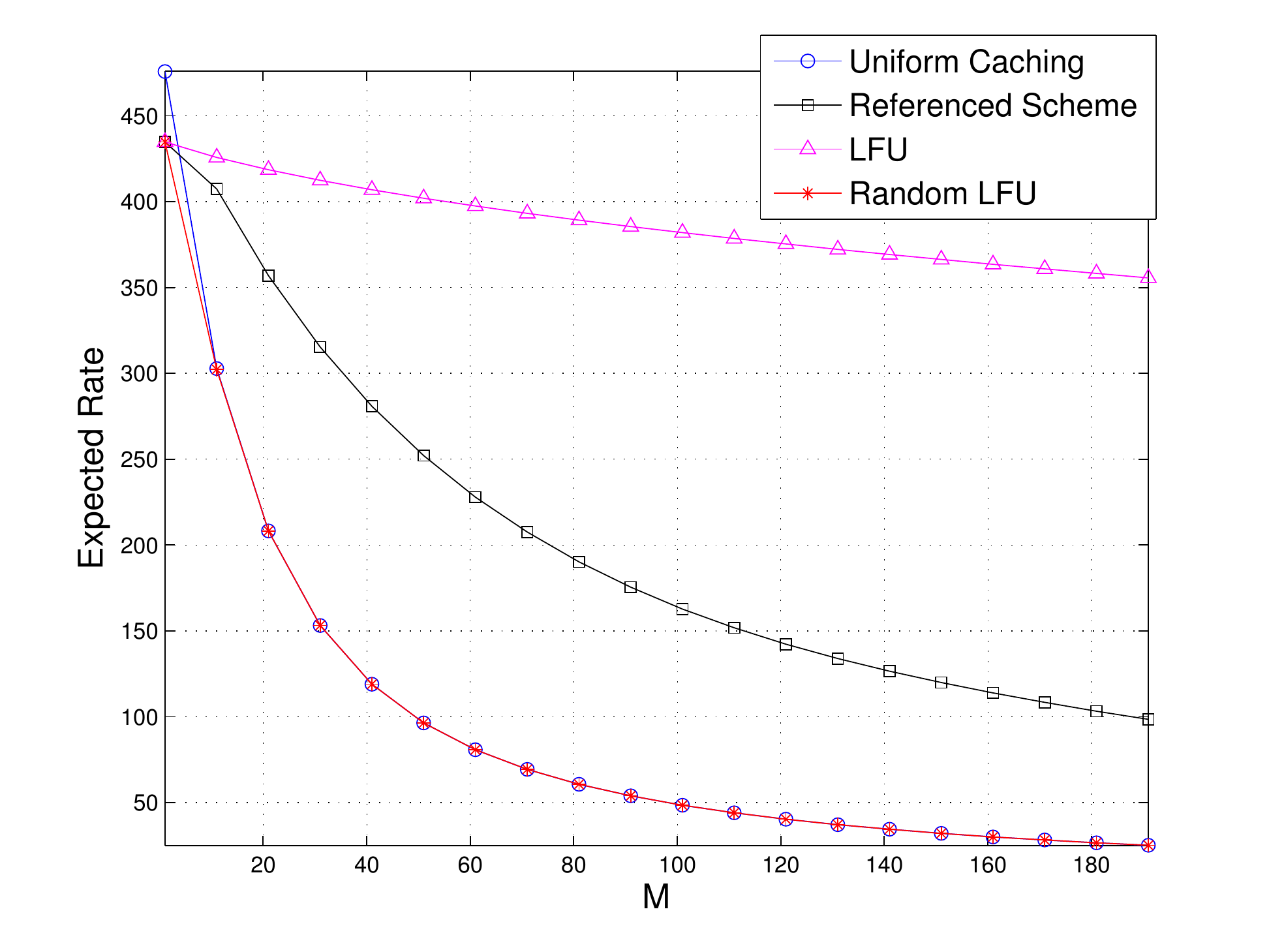}
\label{fig: }
}
\hspace{-0.8cm}
\subfigure[]{
\centering \includegraphics[width=4.7cm]{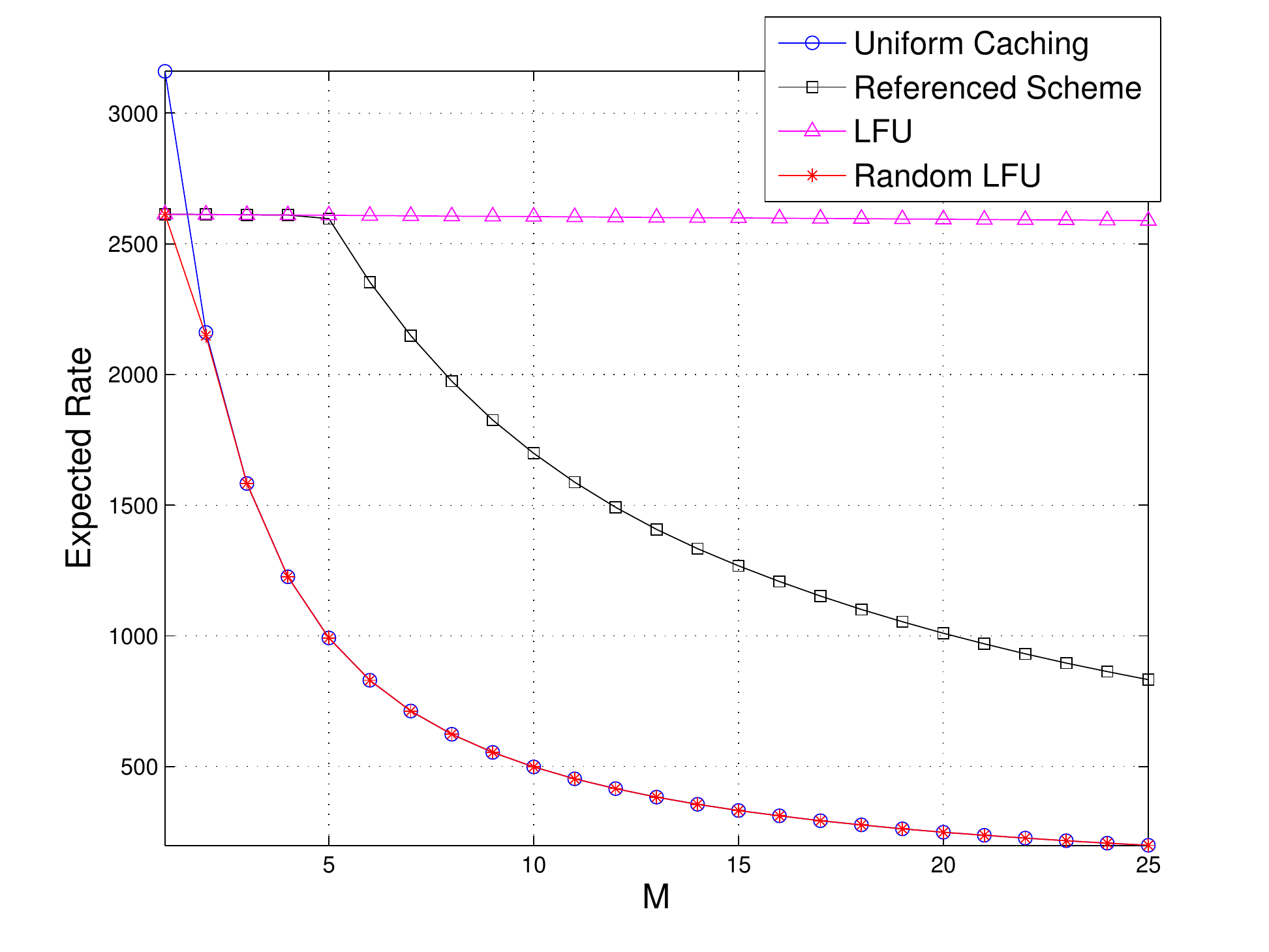}
\label{fig: }
}
\hspace{-0.8cm}
\subfigure[]{
\centering \includegraphics[width=4.7cm]{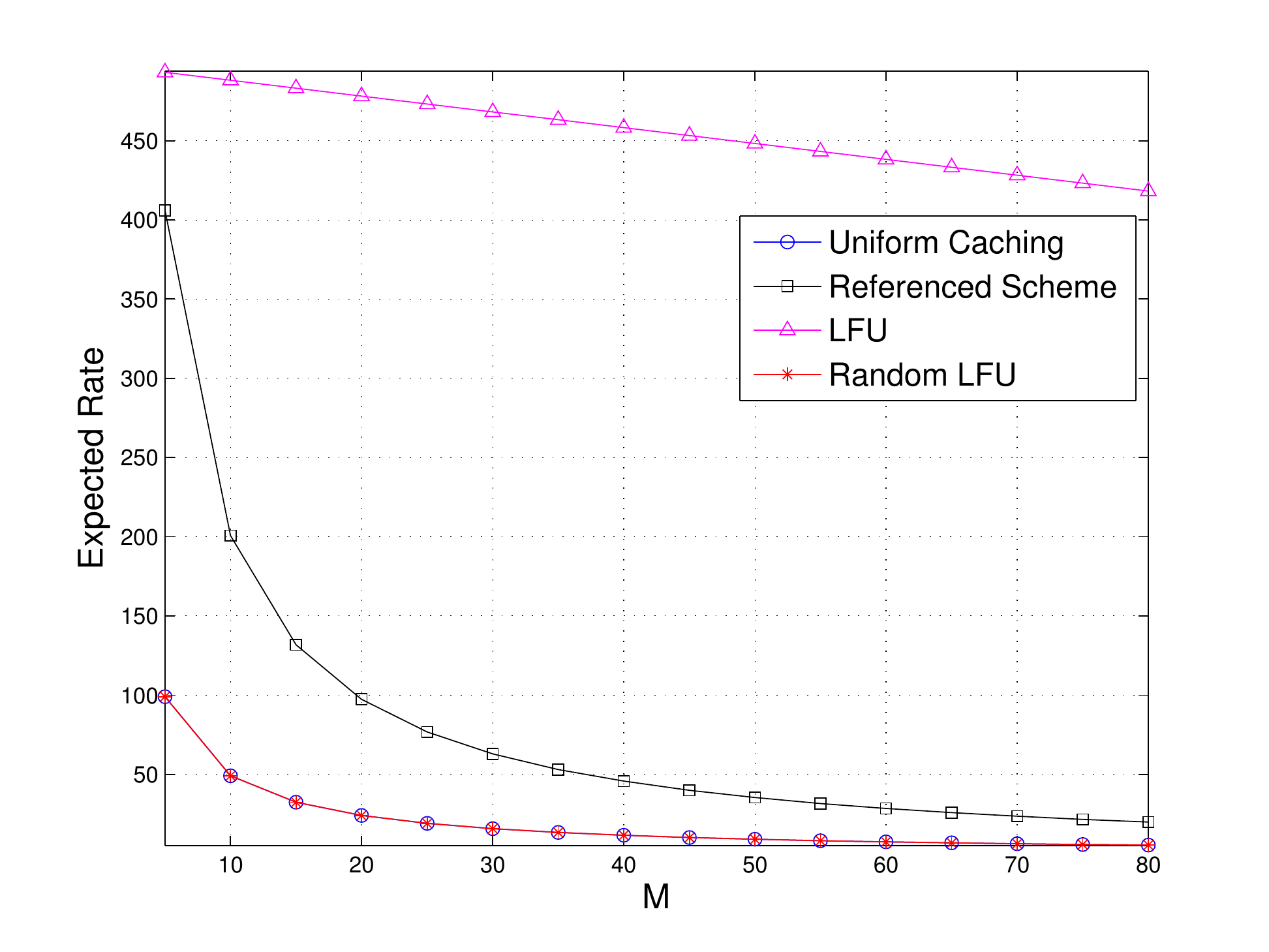}
\label{fig: }
}
\caption{Simulation results for $\alpha=0.6$. a)~$m=5000,n=50$. b)~$m=5000,n=500$. c)~$m=5000,n=5000$. d)~$m=500,n=5000$.}
\label{fig: result 1}
\end{figure*}

\begin{figure*}[ht]
\centering
\subfigure[]{
\centering \includegraphics[,width=4.7cm]{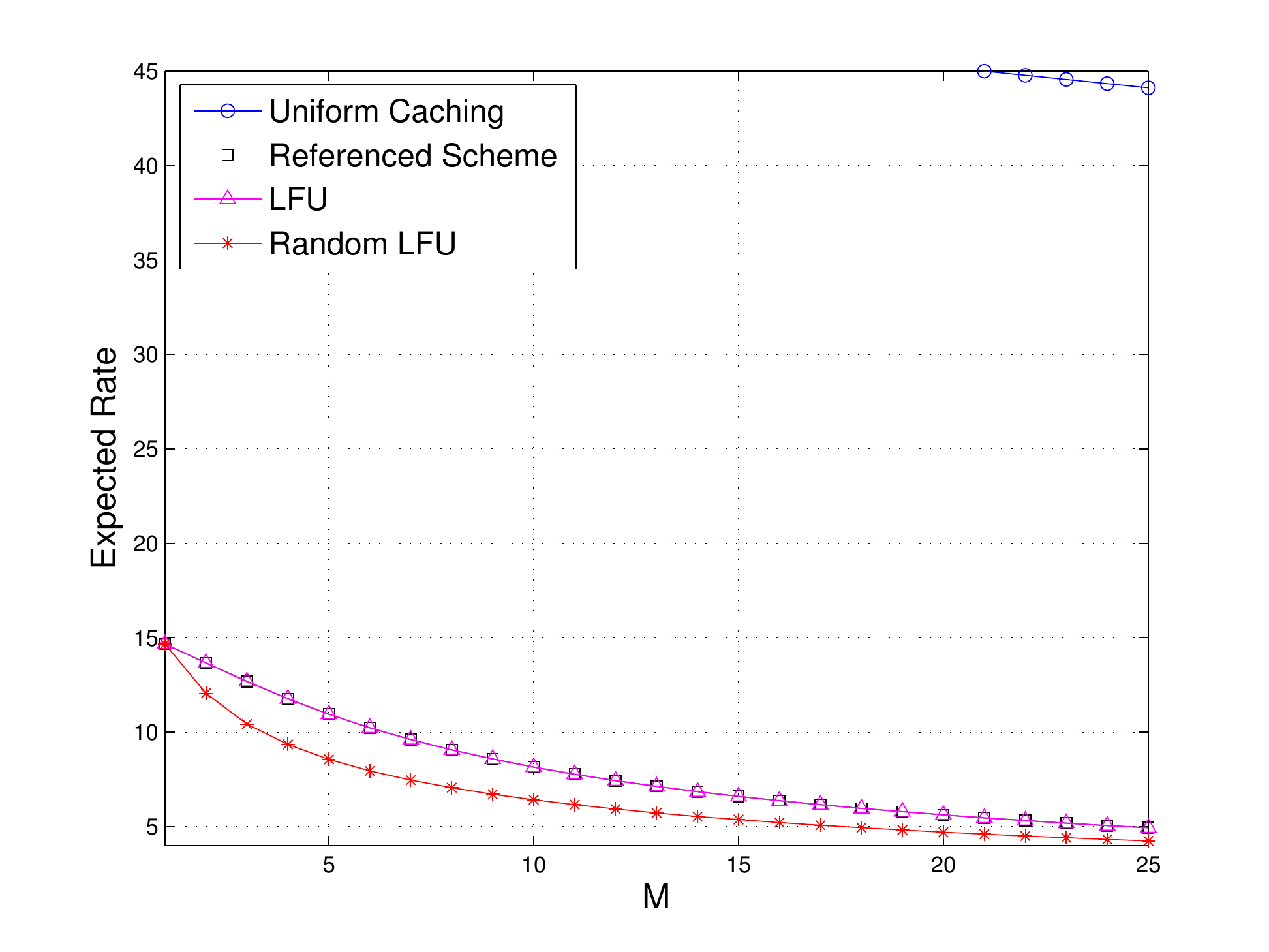}
\label{fig: }
}
\hspace{-0.8cm}
\subfigure[]{
\centering \includegraphics[width=4.7cm]{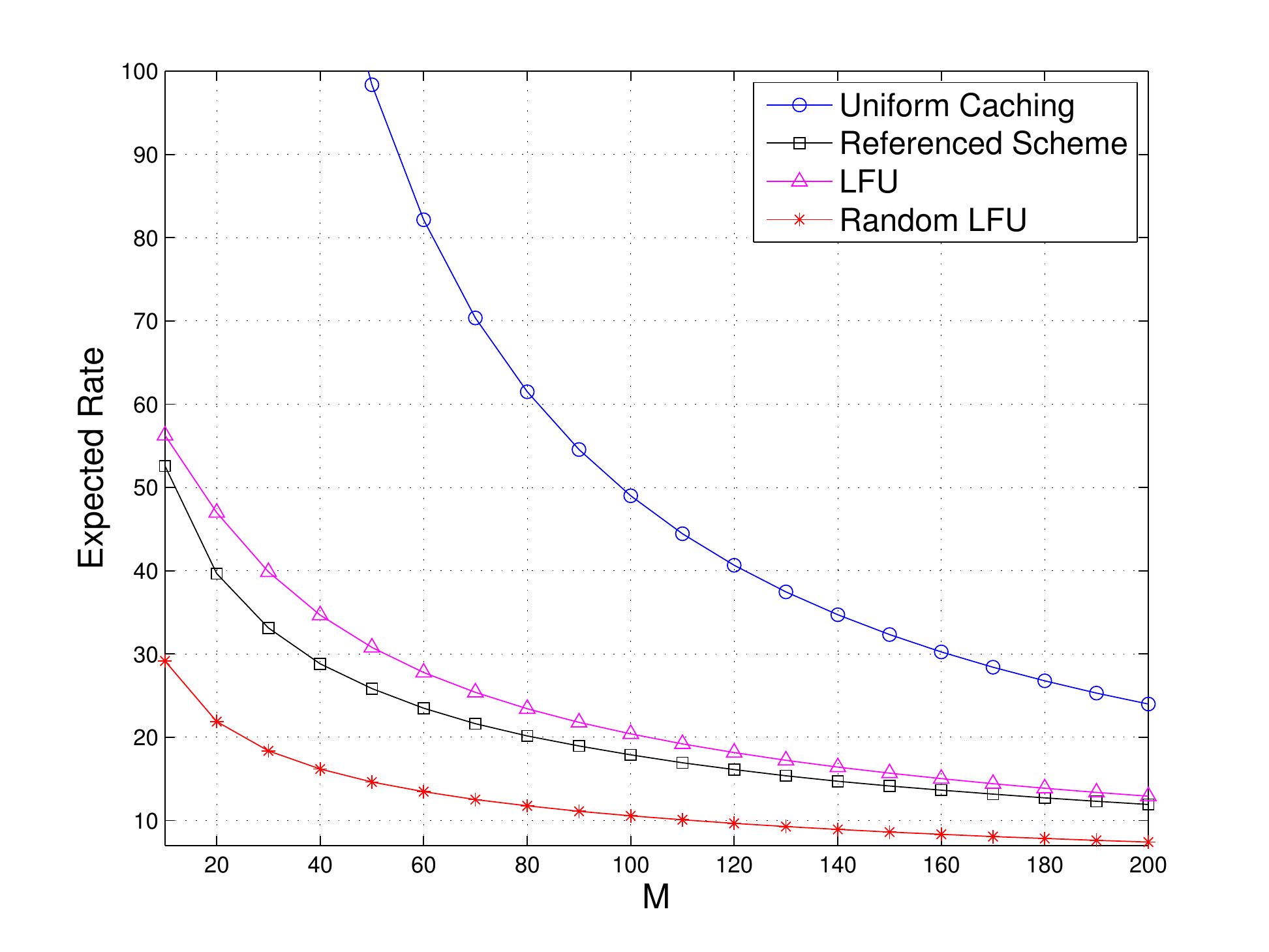}
\label{fig: }
}
\hspace{-0.8cm}
\subfigure[]{
\centering \includegraphics[width=4.7cm]{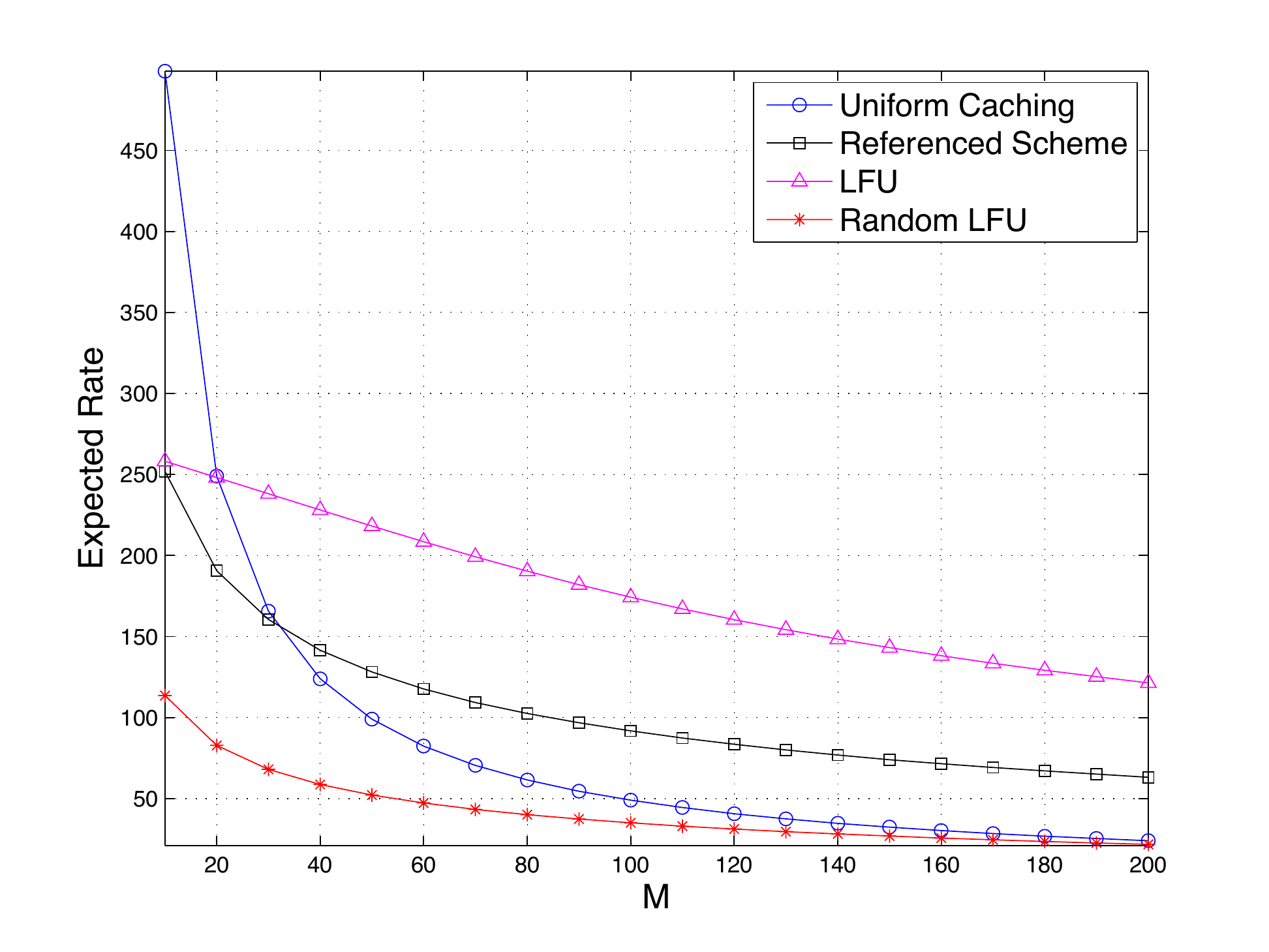}
\label{fig: alpha1p6_m5000_n5000_v5}
}
\hspace{-0.8cm}
\subfigure[]{
\centering \includegraphics[width=4.7cm]{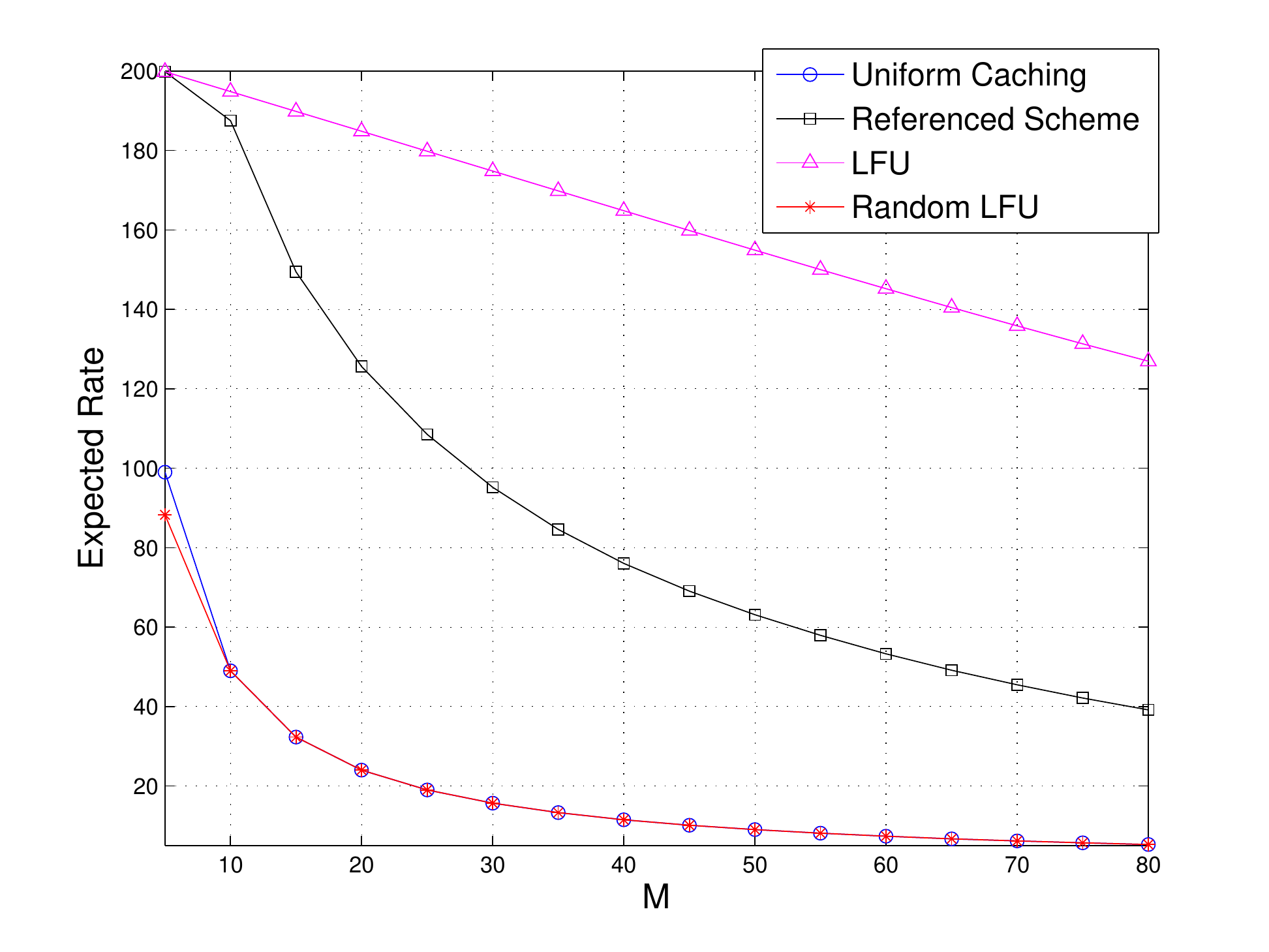}
\label{fig: alpha1p6_m500_n5000_v4}
}
\caption{Simulation results for $\alpha=1.6$. a)~$m=5000,n=50$. b)~$m=5000,n=500$. c)~$m=5000,n=5000$. d)~$m=500,n=5000$.}
\label{fig: result 2}
\end{figure*}

Observe that unlike uncoded delivery schemes that transmit each non-cached packet separately, 
or the scheme proposed in \cite{niesen2013coded}, where files are grouped into subsets and coding is only 
performed within each subset,  the schemes proposed in this paper allow coding over the entire set of requested packets. 
When treating different subsets of files separately,  missed coding opportunities can significantly degrade multicast efficiency. This is shown in  
Fig. \ref{fig: result 1} and \ref{fig: result 2},  where we plot the upper bound of the rate achieved by Random LFU, 
$\bar R^{\rm ub}(\tilde \pv,\qv)$, and compare it with LFU, the scheme proposed in \cite{niesen2013coded} (referenced scheme) and uniform caching.~\footnote{The achievable rate for the scheme proposed in \cite{niesen2013coded} is computed based on a grouping of the files, 
an optimization of the memory assigned to each group, and a separate coded transmission scheme for each group, as described in \cite{niesen2013coded}.} 
The expected rate is shown as a function of the per user cache capacity, $M$, for various values of $m,n$. 

In practice and also in our simulations, $\bar R^{\rm ub}(\tilde \pv,\qv)$ is computed with $\tilde \pv$ in \eqref{ptilde} 
and $\tilde m = \arg \! \min \bar R^{\rm ub}(\tilde \pv,\qv)$. Notice that this optimization problem, which can be solved by simply one dimensional 
search,  is much simpler than the non-convex optimization problem 
needed to optimize the referenced scheme in \cite{niesen2013coded}. 

The simulation results agree with the analytical study shown in Section \ref{order opt}, which is generally difficult to evaluate and verify for the study of scaling laws.  
Observe that for all scenarios, the proposed scheme is able to significantly outperform both LFU and the referenced scheme unless 
the proposed scheme converges to LFU.  
In particular, when $\alpha=1.6$, $m=500$ and $n=5000$, observe from Fig. \ref{fig: alpha1p6_m500_n5000_v4} that for cache size $4\%$ of the library size ($M=20$), the proposed scheme achieves a factor improvement in expected rate of $5\times$ with respect to the referenced scheme and $8\times$ with respect to LFU. 

We remark that the improved performance and order optimality guarantees of the proposed schemes are based on the ability to 1) cache more packets of the more popular files, 2) maximize the amount of distinct packets of each file collectively cached, and 3) allow coded multicast transmissions within the full set of requested packets.


\bibliographystyle{IEEEbib}
\bibliography{references,references_d2d}

\end{document}

%% file: macros.tex
\long\def\comment#1{}

\newfont{\bbb}{msbm10 scaled 700}

\newfont{\bb}{msbm10 scaled 1100}

\newcommand{\PP}{\mbox{\bb P}}

\newcommand{\FF}{\mbox{\bb F}}

\newcommand{\EE}{\mbox{\bb E}}

\newcommand{\pv}{{\bf p}}
\newcommand{\qv}{{\bf q}}

\newcommand{\Fc}{{\cal F}}

\newcommand{\Uc}{{\cal U}}

\newcommand{\Hsf}{{\sf H}}

\newcommand{\Msf}{{\sf M}}

\newcommand{\Qsf}{{\sf Q}}

\renewcommand{\arg}{{\hbox{arg}}}

\newcommand{\eqdef}{\stackrel{\Delta}{=}}

\newcommand{\be}{\begin{equation}}
\newcommand{\ee}{\end{equation}}
\newcommand{\bea}{\begin{eqnarray}}
\newcommand{\eea}{\end{eqnarray}}

